\documentclass[preprint]{aastex62}
\usepackage{xcolor, soul, rotating, graphicx}
\hypersetup{linkcolor=red,citecolor=blue,filecolor=cyan,urlcolor=magenta}

\shorttitle{Ca II H Jets}
\shortauthors{Cho et al.}

\begin{document}
\title{A New Type of Jets in a Polar Limb of Solar Coronal Hole}

\correspondingauthor{Yong-Jae Moon}
\email{moonyj@khu.ac.kr}

\author[0000-0001-7514-8171]{Il-Hyun Cho}
\affiliation{Department of Astronomy and Space Science, Kyung Hee University, Yongin, 446-701, Korea}

\author[0000-0001-6216-6944]{Yong-Jae Moon}
\affiliation{Department of Astronomy and Space Science, Kyung Hee University, Yongin, 446-701, Korea}
\affiliation{School of Space Research, Kyung Hee University, Yongin, 446-701, Korea}

\author[0000-0003-2161-9606]{Kyung-Suk Cho}
\affiliation{Space Science Division, Korea Astronomy and Space Science Institute, Daejeon 305-348, Korea}
\affiliation{Department of Astronomy and Space Science, University of Science and Technology, Daejeon 305-348, Korea}

\author[0000-0001-6423-8286]{Valery M. Nakariakov}
\affiliation{School of Space Research, Kyung Hee University, Yongin, 446-701, Korea}
\affiliation{Centre for Fusion, Space and Astrophysics, Department of Physics, University of Warwick, CV4 7AL, UK}
\affiliation{Special Astrophysical Observatory, Russian Academy of Sciences, St. Petersburg, 196140, Russia}

\author[0000-0001-6412-5556]{Jin-Yi Lee}
\affiliation{Department of Astronomy and Space Science, Kyung Hee University, Yongin, 446-701, Korea}

\author{Yeon-Han Kim}
\affiliation{Space Science Division, Korea Astronomy and Space Science Institute, Daejeon 305-348, Korea}

\begin{abstract}
A new type of chromospheric jets in a polar limb of a coronal hole is discovered
in the Ca II filtergram of the Solar Optical Telescope on board the \textit{Hinode}.
We identify 30 jets in the Ca II movie of duration of 53 min.
The average speed at their maximum heights is found to be 132$\pm$44 km s$^{-1}$
ranging from 57 km s$^{-1}$ to 264 km s$^{-1}$ along the propagation direction.
The average lifetime is 20$\pm$6 ranging from 11 seconds to 36 seconds.
The speed and lifetime of the jets are located at end-tails of those parameters determined for type II spicules,
hence implying a new type of jets.
To confirm whether these jets are different from conventional spicules,
we construct a time-height image averaged over horizontal region of 1$\arcsec$,
and calculate lagged cross-correlations of intensity profiles at each height with the intensity at 2 Mm.
From this, we obtain a cross-correlation map as a function of lag and height.
We find that the correlation curve as a function of lag time is well fitted into three different Gaussian functions
whose standard deviations of the lag time are 193 seconds, 42 seconds, and 17 seconds.
The corresponding propagation speeds are calculated to be 9 km s$^{-1}$, 67 km s$^{-1}$, and 121 km s$^{-1}$, respectively.
The kinematic properties of the former two components seem to correspond to the 3 minutes oscillations and type II spicules,
while the latter component to the jets addressed in this study.
\end{abstract}

\keywords{Solar chromosphere (1479), Astronomy data analysis (1858), Solar optical telescopes (1514)}

\section{Introduction} \label{sec:intro}
The solar chromosphere is morphologically dominated by spicular jets.
It is suggested that there are at least two types of spicules when observed in the Ca II filtergram \citep{2007PASJ...59S.655D}.
One is so called type I spicules typically observed in active regions.
They show a rise and fall,
have lifetimes of 150 -- 400 seconds and maximum ascending velocities of 15 -- 40 km s$^{-1}$ \citep{2012ApJ...759...18P}.
Their motion is characterized by a non-ballistic and parabolic-like path \citep{2007PASJ...59S.655D},
suggesting that they are driven by shocks \citep{2007ASPC..368...65D},
which could be observed in fibrils and mottles \citep{1995ApJ...450..411S, 2006ApJ...647L..73H, 2007ApJ...660L.169R}.

The other is type II spicules, which is fast and short lived comparing to type I spicules \citep{2007PASJ...59S.655D}.
It was shown that type II spicules, common in the quiet Sun and coronal holes,
have lifetimes of 50 -- 150 seconds, velocities of 30 -- 110 km s$^{-1}$,
and are not seen to fall down, but fade at around their maximum length \citep{2012ApJ...759...18P}.
On the other hand, \citet{2012ApJ...750...16Z} reported that there seems no convincing evidence of type-II spicules,
and the majority of spicules is type-I spicules.
Hence, there is still an ongoing debate about the existence of type-II spicules.

\citet{2014Sci...346A.315T} discovered intermittent small-scale jets in narrow network lanes
in the transition region lines with the Interface Region Imaging Spectrograph \citep[IRIS,][]{2014SoPh..289.2733D}.
The average speed, lifetime, recurring period of the network jets were found to be
80 -- 250 km s$^{-1}$, 20 -- 80 seconds, and 8.3 minutes, respectively, suggesting they could be a counterpart of Type II spicule.
It is proposed that some of the network jets are propagating heating fronts caused by dissipation of a current-sheet
that could be formed during a formation of the spicule \citep{2017ApJ...849L...7D}.
From multi-wavelength observations in Ca II H with the Solar Optical Telescope \citep[SOT,][]{2008SoPh..249..167T}
on board the \textit{Hinode} \citep{2007SoPh..243....3K}, IRIS ultraviolet (UV)
and extreme ultraviolet (EUV) filtergrams taken by the Atmospheric Imaging Assembly \citep[AIA,][]{2012SoPh..275...17L}
on board the \textit{Solar Dynamics Observatory} \citep[\textit{SDO},][]{2012SoPh..275....3P},
it was shown that Ca II spicules often follow only the early, more linear, phase of the parabola and
mostly continue their evolution in hotter pass bands after they fade from Ca II H line \citep{2015ApJ...806..170S}.
\citet{2019ApJ...871..230L} analyzed 330 macrospicules with the lifetime of $\sim$10 minutes
observed in the 304 {\AA} of the \textit{SDO}/AIA.
In their study, a part of macrospicules was found to be inconsistent with a purely ballistic motion as in the type I spicules.
But the rise velocity is 70--140 km s$^{-1}$, which is similar to the speed of type II spicules.
One the other hand, \citet{2012ApJ...759...18P} discovered faint and transient spicules in active regions,
whose speed and lifetime are 160 $\pm$ 36 km s$^{-1}$ and 37 $\pm$ 16 seconds, respectively.
These were interpreted as type II spicules in active regions based on their high velocities and short lifetimes.

In this Letter,
we report on a new type of highly transient and recurring jets in a polar coronal hole observed in the Ca II H filtergram.
In section 2, we present data processing.
In section 3, kinematic properties of jets are presented by two different methods.
Finally, we summarize and discuss our results.

\section{Data} \label{sec:data}
We use the Ca II H filtergrams taken by the \textit{Hinode}/SOT on 2011-Jan-29 11:06 -- 11:59 UT.
The time cadence, pixel resolution, and exposure time are 1.6 s, 0.109$\arcsec$ , and $\sim$0.6144 seconds, respectively.
The wavelength passband of the filtergram is 3968.5$\pm$3 {\AA}.
The field-of-view is 56$\arcsec$$\times$56$\arcsec$ which is located at the south polar coronal hole.
This data has been analyzed by \citet{2011ApJ...736L..24O} to explore propagating kink waves
along type II spicules lasting for longer than 40 seconds.

\begin{figure}
\centering
\includegraphics[scale=0.7]{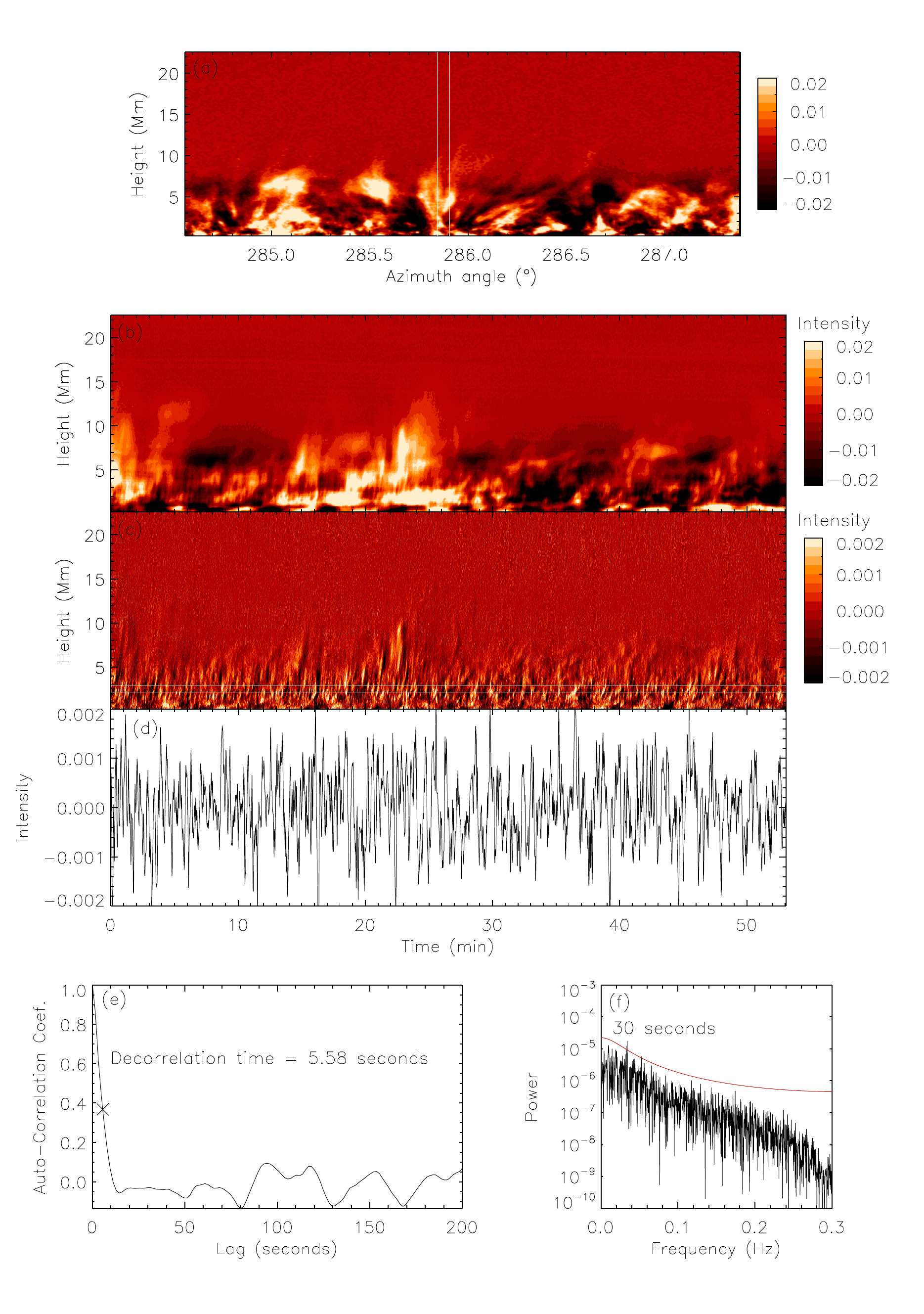}
\caption{
A snapshot of the Ca II filtergram image normalized by an average intensity of the disk region (a),
time-height image for the slice defined by two vertical white lines (b),
first order time-difference of the time-height image (c),
average intensity time series of the region between two horizontal white lines, which is subtracted by its linear least-square fitting line (d),
autocorrelation function of the de-trended intensity time series with its decorrelation time ($\times$) (e),
and corresponding Fourier power (f).
In panel (a), the radial gradient is subtracted from the original image.
The red curve in panel (f) is the noise power of 99.9\% significance level of the red noise defined by the autoregressive noise model of the order 1.
}
\label{fig1}
\end{figure}

We align the image cube by using the disk region,
and rotate it to locate limbs approximately parallel to the horizontal line.
Then, each image is transformed to the polar coordinate by using bilinear interpolation.
In this step, the horizontal and vertical axes of the polar coordinate are defined by sampling positions evenly
along a straight line which is normal to the limb for a given position along the limb.
There is a very tiny offset in the cadence of 1.6 seconds.
We use a 1.6 seconds evenly sampled image cube by using the linear interpolation.
The intensity is normalized by the mean intensity of the solar disk for a given frame.
For a given height, the radial gradient determined from the average intensity in the horizontal direction is subtracted (Figure~\ref{fig1}a).

\section{Methods and Results} \label{sec:mr}
From the background subtracted (radial gradient) Ca II movie,
we recognized that there are very transient, repeated and fast jet-like flows or eruptions.
To identify what they are, we draw the time-height image over a 1$\arcsec$ horizontal region indicated by two vertical white lines in Figure~\ref{fig1}a.
As shown in Figure~\ref{fig1}b, there seems to exist many repeated, thin, faint, and inclined straws in the time-height image.
To quantitatively verify if the straw structures are real,
we estimate the Fourier power for the time series of de-trended intensity (Figure~\ref{fig1}d).
The de-trended intensity is defined as the average of the first order time-differences over the height range of 2--3 Mm
indicated by two horizontal white lines in Figure~\ref{fig1}c, and then subtracted by its linear least-square fitting line.
The Fourier power seems to be inversely proportional to frequency, and hence we interpret that the time series includes a red-noise.
The noise power is explicitly given by the autoregressive model of the order 1 \citep{2002CG.....28..421S}
as $P = \sigma^2 / (1 - 2\rho \cos(\pi f/f_N) + \rho^2)$, which follows a chi-square distribution with two degrees of freedom,
where $\sigma$ and $\rho$ are the standard deviation of the time series and the autoregressive parameter, respectively.
The autoregressive parameter is determined from the decorrelation time ($\tau$) of the autocorrelation function of the time series, $\rho = e^{-1.6/\tau}$,
where 1.6 (seconds) indicates the sampling interval of the data.
The decorrelation time is defined as the time when the autocorrelation is equal to $e^{-1}$ ($\times$ in Figure~\ref{fig1}e).
This approach provides noise powers resembling the power-law \citep[e.g.,][]{2017A&A...602A..47P}
where the period is close to the decorrelation time, but provides uniform-like powers as the period locates further away from the decorrelation time.
It is found that the powers at frequency equal to 1/(30 seconds) are apparently higher than the noise level of 99.9\% (red curve).
The periodicity of 30 seconds could indicate that the straws appear at every 30 seconds in the time-height image.

In following subsections, we identify 30 set of jet evolutions in the Ca II movie, which were expected from the inclined straws.
Independently, we identify three upwardly propagating components in the Ca II movie by a cross-correlation analysis.
As the next step, we demonstrate that the kinematics of our jets are different from conventional spicules.

\begin{sidewaysfigure}
\includegraphics[scale=0.8, angle=90]{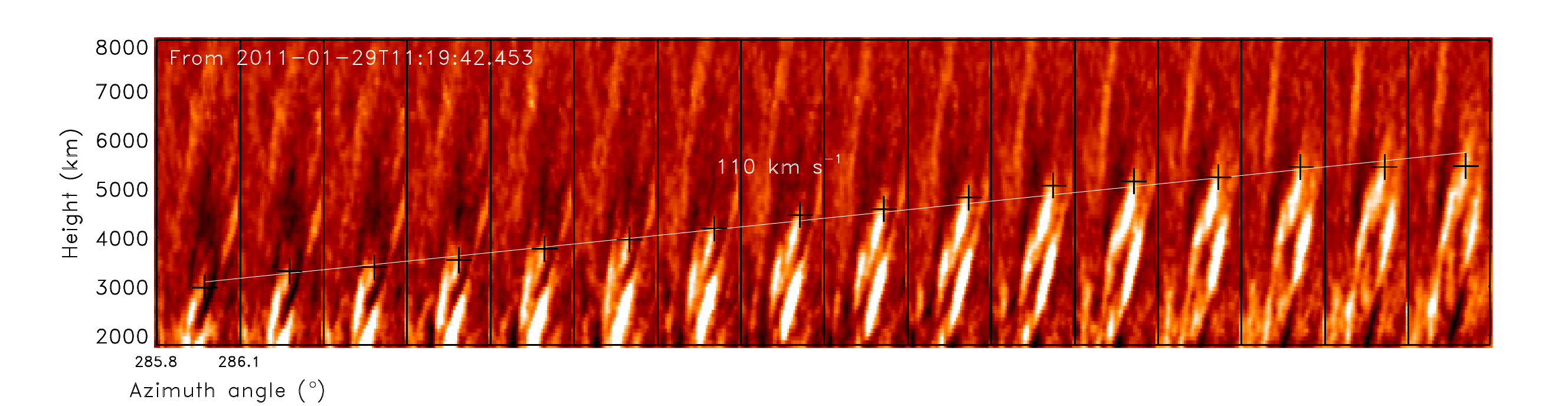}
\includegraphics[scale=0.8, angle=90]{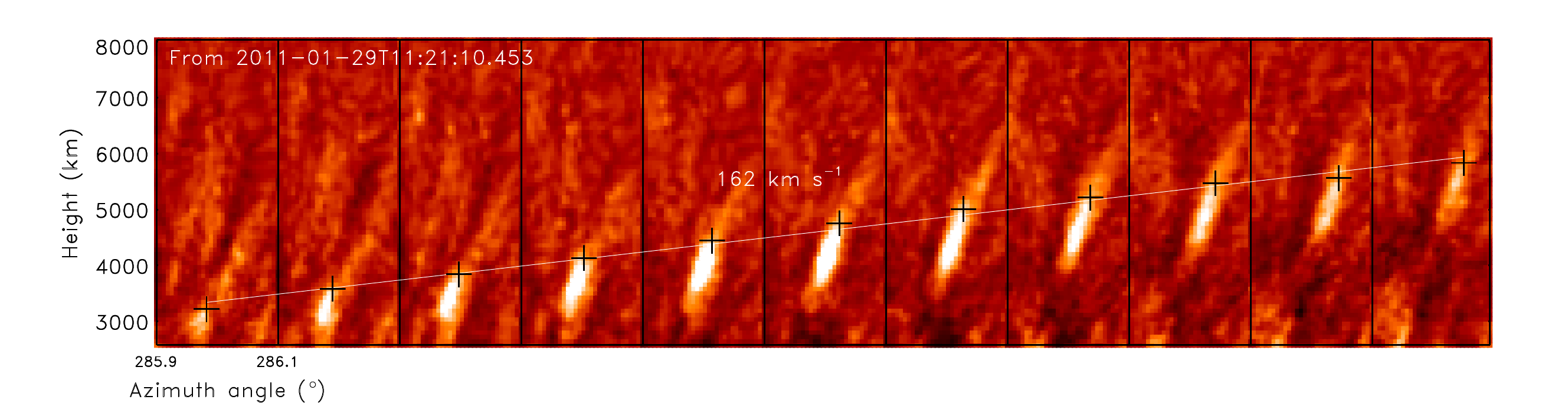}
\includegraphics[scale=0.8, angle=90]{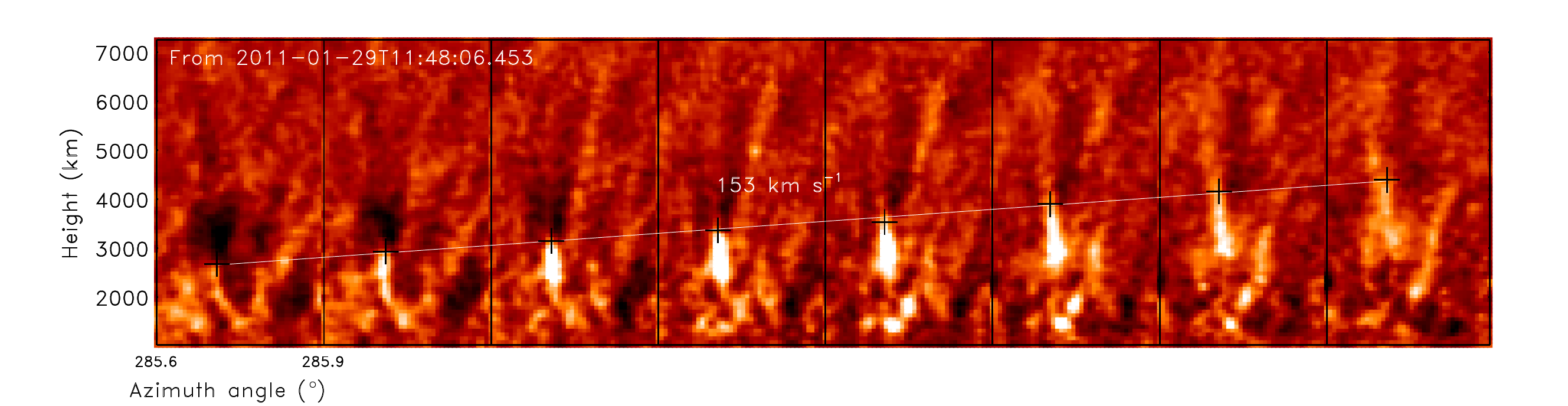}
\caption{
Three examples of the evolution of jets at every 1.6 seconds.
The cross symbol represents the maximum height of a jet at each frame.
The white-straight line is the linear fitting curve of maximum heights, which indicates the speed of jets in vertical direction.
Additional examples are given in Appendix.
}
\label{fig2}
\end{sidewaysfigure}

\subsection{Kinematics of faint jets}
Based on the estimated period of the occurrence, the Ca II intensity is subtracted by backgrounds given by a smoothing window of 30 seconds at each pixel position.
By this process, the temporal variation of type II spicules would be minimized, enhancing a jet-like feature on the plane of the sky,
which was expected from the inclined straws in the time-height image.
From the background subtracted movie, we identify 30 set of jet-like evolutions.
We determine the maximum heights as a function of time for each jet,
and calculate the lifetime as well as the vertical propagation speed of maximum height using a linear least-square fitting.
We also determine the inclination angle and propagation distance of maximum height of each jet.
Three examples are shown in Figure~\ref{fig2} and remaining 27 events are provided in Appendix.

\begin{figure}
\centering
\includegraphics[scale=0.7]{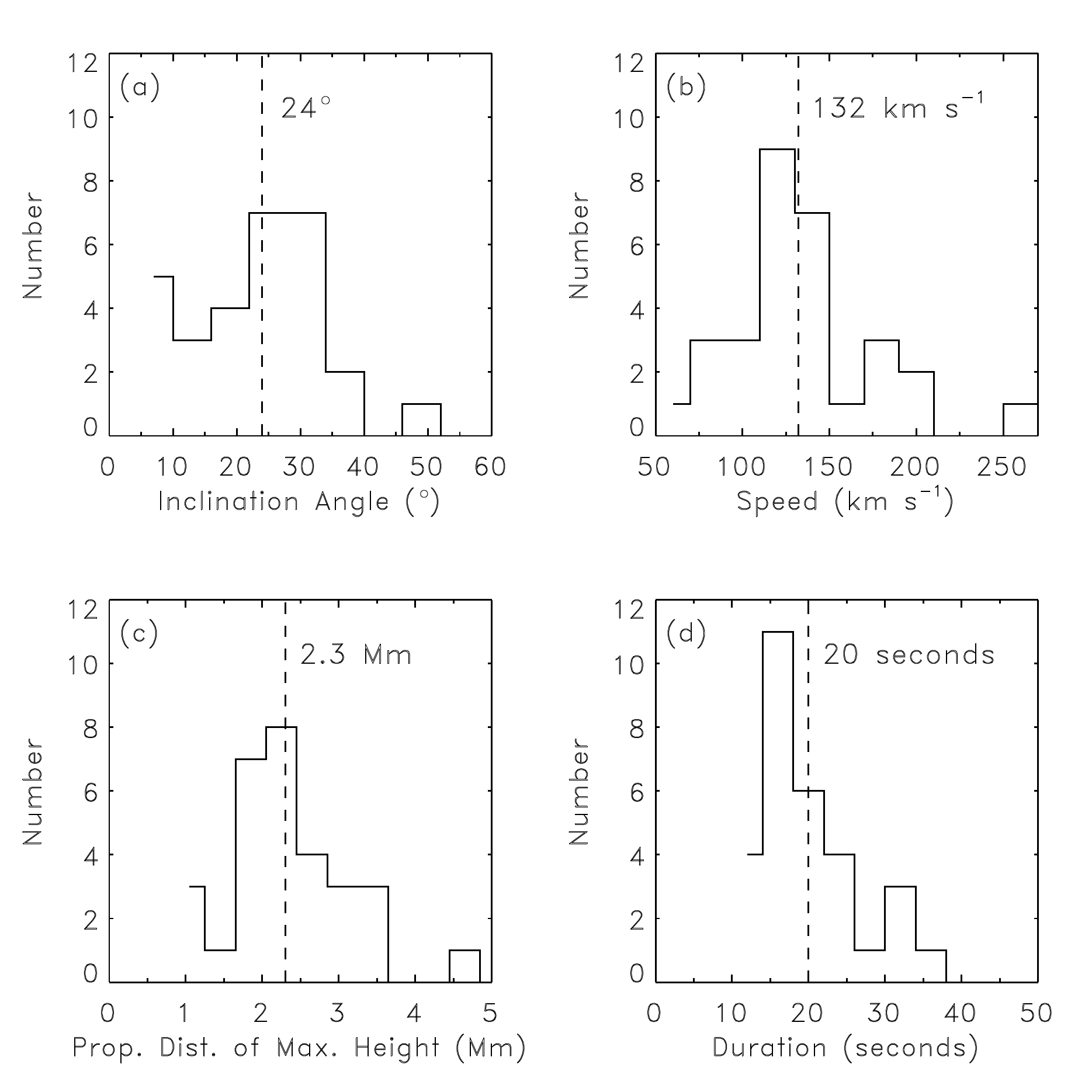}
\caption{
Histograms of inclination angle (a), speed (b), propagation distance (c), and duration (d) of 30 jets.
The vertical dashed line in each panel indicates the mean value of the parameter.
}
\label{fig3}
\end{figure}

The histograms of the jet parameters are presented in Figure~\ref{fig3}.
The inclination angle of the jets is 24$\pm$12$^{\circ}$ and ranges from 4$^{\circ}$ to 53$^{\circ}$.
It is found that the propagation distance of maximum heights of the jets is 2.3$\pm$0.8 Mm ranging in 0.8 -- 4.6 Mm.
The lifetime is 20$\pm$6 seconds ranging in 11 -- 36 seconds.
The speed along the inclined path is calculated to be 132$\pm$44 km s$^{-1}$ and ranges from 57 km s$^{-1}$ to 264 km s$^{-1}$.
The mean speed is located outside two sigma from the average speed of type II spicules,
and the lifetime around two sigma,
indicating that the jets are more transient and faster comparing to those of type II spicules in the Ca II band \citep[c.f.,][]{2012ApJ...759...18P}.
A comparison between the kinematics of the jets discussed here and conventional spicules are presented in Table~\ref{table1}.

\begin{deluxetable*}{cccc}
\tablecaption{Comparison of kinematic parameters of the jets \label{table1}}
\tablehead{
\colhead{} & \colhead{Type I spicules\tablenotemark{a}} & \colhead{Type II spicules\tablenotemark{b}} & \colhead{New jets}
}
\startdata
Inclination Angle ($^{\circ}$) & 12.7$\pm$9.8 & 15.3$\pm$11.1 & 24$\pm$12 \\
Speed (km s$^{-1}$) & 30$\pm$9 & 71$\pm$29 & 132$\pm$44 \\
Lifetime (seconds) & 262$\pm$80 & 83$\pm$35 & 20$\pm$6 \\
Propagation Distance (Mm) & & & 2.3$\pm$0.8 \\
Recurring Period (seconds) & & & 30 \\
\enddata
\tablenotetext{a}{ in active regions given by \citet{2012ApJ...759...18P} }
\tablenotetext{b}{ in coronal holes given by \citet{2012ApJ...759...18P} }
\end{deluxetable*}

\subsection{Cross-correlation analysis of time-height images}
\begin{figure}
\includegraphics[scale=0.7, angle=90]{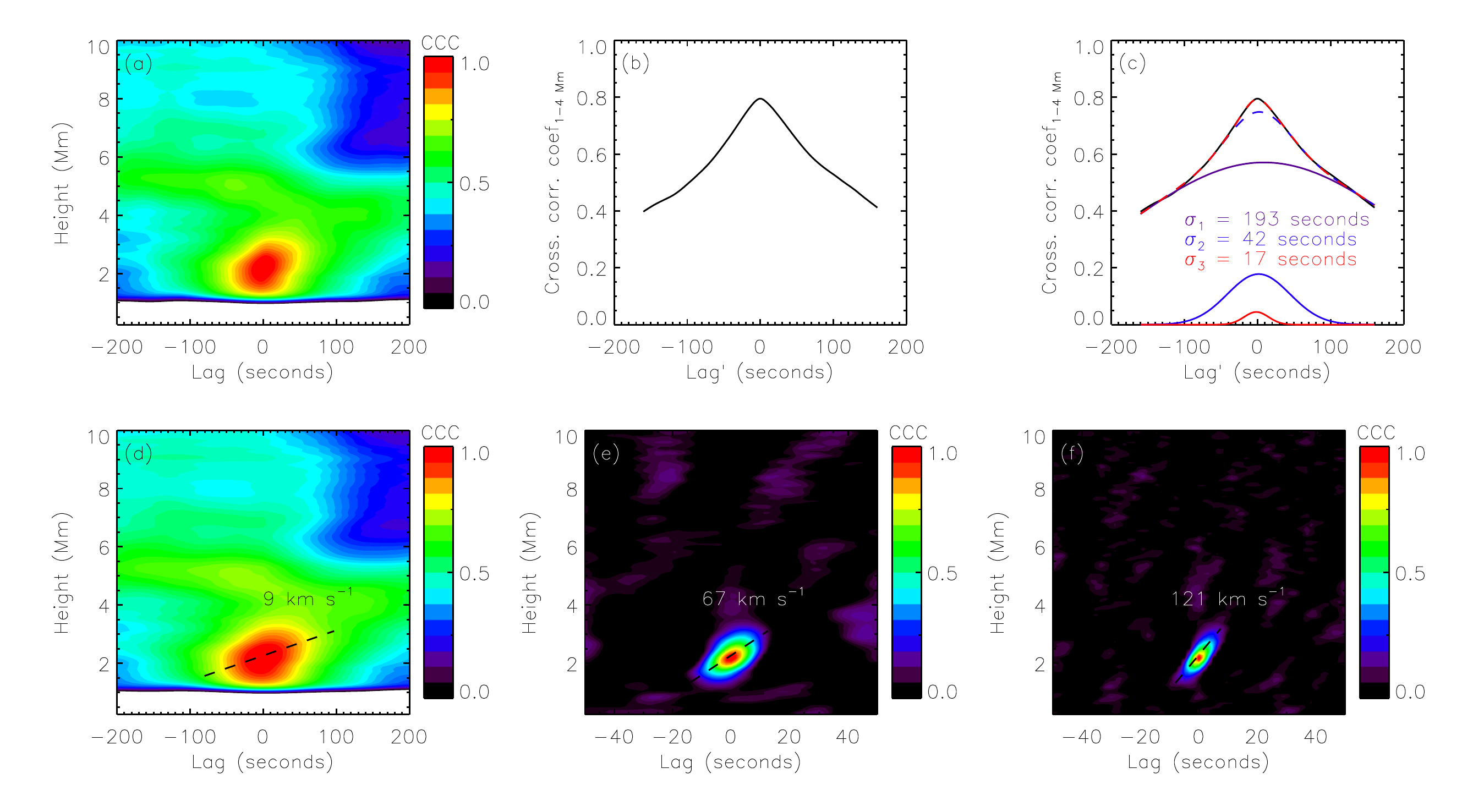}
\caption{
The map of cross-correlation coefficient (CCC) obtained from the time-height image of Figure~\ref{fig1}b (a)
and average of CCC in the height range from 1 to 4 Mm (b).
The lagged-correlation at each height is calculated from two intensity profiles at the given height and 2 Mm.
In panel (c), the purple, blue and red curves are three Gaussian functions with the standard deviations of 193, 42, and 17 seconds,
respectively.
The blue dashed curve is the sum of the former two Gaussian functions.
The red dashed curve is the sum of all the three functions.
The cross-correlation map in panel (d) is obtained from the time-height images of smoothed backgrounds whose width of averaging windows is 193 seconds.
The cross-correlation maps in panel (e) and (f) are obtained from the time-height images
subtracted by smoothed backgrounds whose widths of averaging windows are 42, and 17 seconds.
The straight-dashed line in panel (d)--(f) indicates the propagation speed of intensity variations in the time-height images.
}
\label{fig4}
\end{figure}

As mentioned above, the kinematic properties of our jets seem to be located at an high-end tail of the distribution of the II spicules.
To examine whether these jets are different from type II spicules,
we perform a lagged cross-correlation analysis to identify propagating components of intensity disturbances independently.
From the original time-height image shown in Figure~\ref{fig1}b,
we calculate lagged cross-correlation coefficient (CCC) between two intensity profiles across heights (Figure~\ref{fig4}a).
One is the intensity profile for a given height, and the other is at a fixed height of 2 Mm.
The lag is set from +320 seconds to -320 seconds with 1.6 seconds interval.
We average the correlation coefficients from 1 to 4 Mm as shown in Figure~\ref{fig4}b, with respect to the zero lag for a given height.
With this analysis, we expect that at least two propagating components of intensity disturbances would be captured in the correlation map,
including type II spicules and our jets.

It is found that the curve is well fitted with a sum of three Gaussian functions (rather than two components)
with standard deviations of 193, 42, and 17 seconds,
which are indicated by solid-purple, solid-blue, and solid-red curves, respectively (Figure~\ref{fig4}c).
In Figure~\ref{fig4}c, the dashed-blue and dashed-red curves represent the sum of former two Gaussian functions
and the sum of all the three Gaussian functions, respectively.
Note that the solid black in Figure~\ref{fig4}c is the same with Figure~\ref{fig4}b.

The above analysis may imply that the original time-height image includes three characteristic timescales of different dynamical components.
To identify each component in a cross-correlation map, we construct three different time-height images based on the three timescales.
One image is obtained from the background of the original time-height image, whose timescale is longer than 193 seconds.
The other two images are background subtracted ones, whose timescales are shorter than 42 seconds and 17 seconds, respectively.
From these three filtered time-height images,
we calculate again the cross-correlations as a function of lag and height referenced by the intensity profile at the fixed height of 2 Mm (Figure~\ref{fig4}d--\ref{fig4}f).
The positive lag for a given height indicates that the intensity profile is lagged to the intensity profile at 2 Mm.
The negitive lag indicates that the intensity profile leads to the profile at 2 Mm.
Hence, the inclined elliptic contours shown in the correlation maps in Figure~\ref{fig4}d--\ref{fig4}f represent upward propagations.

The propagation speeds are various in three components.
In Figure~\ref{fig4}d, the speed is calculated as 9 km s$^{-1}$.
Interestingly, this speed corresponds to the sound speed of $\sim$7635 K plasma,
that approximates the propagation speed of slow magnetoacoustic waves under a low-$\beta$ condition \citep{2006RSPTA.364..447R}.
This implies that the speed reflects cool chromospheric conditions.
In addition, the timescale of 193 seconds is corresponding to the period of chromospheric oscillations.
In Figure~\ref{fig4}e, the speed is estimated to be 67 km s$^{-1}$.
Considering the average inclination angle of 24$^{\circ}$, the de-projected speed is 73 km s$^{-1}$
which is well consistent with the typical speed of type II spicules in coronal holes (72 km s$^{-1}$) \citep{2012ApJ...759...18P}.
The timescale of 42 seconds corresponds to the half of the lifetime (83$\pm$35 seconds) of type II spicules in coronal holes \citep{2012ApJ...759...18P},
we interpret that the second component may reflect the kinematic properies of type II spicules too.

However, the third component gives the speed of 121 km s$^{-1}$ (Figure~\ref{fig4}f).
Considering the inclination angle, the de-projected speed becomes 132.5 km s$^{-1}$.
Interestingly this value is the same with the speed of jets in our study.
In addition, the timescale of 17 seconds is also well consistent with the lifetime of the jets (20 seconds).
Hence, the jets presented in our study are likely belonging to the third component which is independent of type II spicules.

\section{Summary and Discussion} \label{sec:sd}
We provide 30 set of jet evolutions in the Ca II filtergram taken by \textit{Hinode}/SOT observed from 2011-Jan-29 11:06 -- 11:59 UT.
The occurrence period, propagation speed, and lifetime are 30 seconds, 132$\pm$44 km s$^{-1}$ (57 -- 264 km s$^{-1}$), and 20$\pm$6 seconds (11 -- 36 seconds), respectively.
The intensity perturbation having similar kinematic properties is detected from the cross-correlation analysis,
which is significantly different from the chromospheric oscillations and type II spicules.
Hence, we conclude that there exist a new type of jet-like propagating structures in the chromosphere, which is more transient and faster than typical type II spicules.
The intensity is at least 10 times fainter than the typical intensity of the chromospheric structure in the limb as shown in Figure~\ref{fig1}b--\ref{fig1}c,
but it seems to occur very frequently (30 seconds), suggesting that these jets may play a role on the mass supply to the solar wind.
On the other hand, the high speeds of these jets are likely not completely caused by mass flows.
A significant fraction could be caused by propagating heating fronts \citep[e.g.,][]{2017ApJ...849L...7D} or fast-mode wave \citep[e.g.,][]{2016ApJ...832..141K}.
The propagation distance of maximum height of the jets is found to be 2.3$\pm$0.8 Mm,
suggesting that these jet might not directly contribute to the coronal heating.

Recently, rapid and transient phenomena were reported in the transition region \citep[e.g.,][]{2014Sci...346A.315T, 2016ApJ...832..141K, 2015ApJ...811L..33V} 
as well as in the corona \citep[e.g.,][]{2013ApJ...770L...1T, 2014ApJ...787..118R, 2015ApJ...809...82G, 2016SoPh..291..155S, 2017Ap&SS.362...10J, 2018ApJ...868L..27P, 2018ApJ...862...18D}.
Particularly, \citet{2014Sci...346A.315T} discovered intermittent small-scale jets in narrow netwrok lanes in the IRIS transition region lines.
The speed, lifetime, and recurring period of the network jets were found to be 80 -- 250 km s$^{-1}$, 20 -- 80 seconds, and 8.3 minutes, respectively.
The speed and lifetime of the jets in our study are overlapped with those of the network jets.
\citet{2016ApJ...832..141K} discovered ubiquitous fast-propagating intensity disturbances (PIDs) on the Sun by the Ly$\alpha$ transition region line
taken by the Chromospheric Lyman-Alpha SpectroPolarimeter \citep[CLASP,][]{2012SPIE.8443E..4FK, 2012ASPC..456..233K}.
The speed and timescale of the PIDs were found to be 150 -- 350 km s$^{-1}$ and 30 seconds, respectively,
which are overlapped with the jet parameters in this study.
Hence, kinematic properties of the Ca II jets in our study may partly be related to the previous observations in the UV bands.
On the other hand, the jets in this study is obviously faster than the speeds of chromospheric jets
\citep[e.g.,][]{2007Sci...318.1591S, 2013SoPh..288...39Y, 2019ApJ...877L...1C}
which are apparently associated with magnetic reconnections \citep[e.g.,][]{2015ApJ...810...38K, 2016ApJ...824...96T, 2019ApJ...871..125S}.

A mechanism responsible for the observed jet evolutions could be a super-Alfv\'{e}nic flow.
A ponderomotive force in a non-linear torsional Alfv\'{e}n wave \citep{2011A&A...526A..80V, 2018ApJ...869...93M},
which could be generated by magnetic tornadoes observed from the photosphere to the corona \citep{2012Natur.486..505W},
can produce a longitudinal (directed upwards) perturbation at a super-Alfv\'{e}nic speed.
On the other hand, high-frequency oscillations
\citep[e.g.,][]{2007A&A...473..943J, 2007A&A...474..627Z, 2012NatCo...3.1315M, 2017ApJS..229....9J, 2019FrASS...6...36N}
and/or periodic reconnection induced by magnetohydrodynamic waves
\citep[e.g.,][]{2012A&A...548A..98M, 2016SSRv..200...75N, 2017ApJ...844....2T, 2018SSRv..214...45M, 2019ApJ...874L..27X}
could be considered responsible for the observed jet occurrence too.

\acknowledgments
This work is supported by the Korea Astronomy and Space Science Institute
under the R\&D program 'Development of a Solar Coronagraph on International Space Station,
the National Research Foundation of Korea (NRF) (grant No. 2019R1C1C1006033, NRF-2019R1A2C1002634, NRF-2016R1A2B4013131, NRF-2016R1A6A3A11932534),
and the BK 21 plus program funded by the Korean Government.
This work is also supported by the Institute for Information \& Communications Technology Promotion (grant No. 2018-0-01422).
V.M.N. acknowledges support from the STFC consolidated grant ST/P000320/1, the Russian Foundation for Basic Research (grant No. 18-29-21016),
and the BK 21 plus program.

\appendix
\section{Additional examples of jets}

\begin{sidewaysfigure}
\includegraphics[scale=0.8, angle=90]{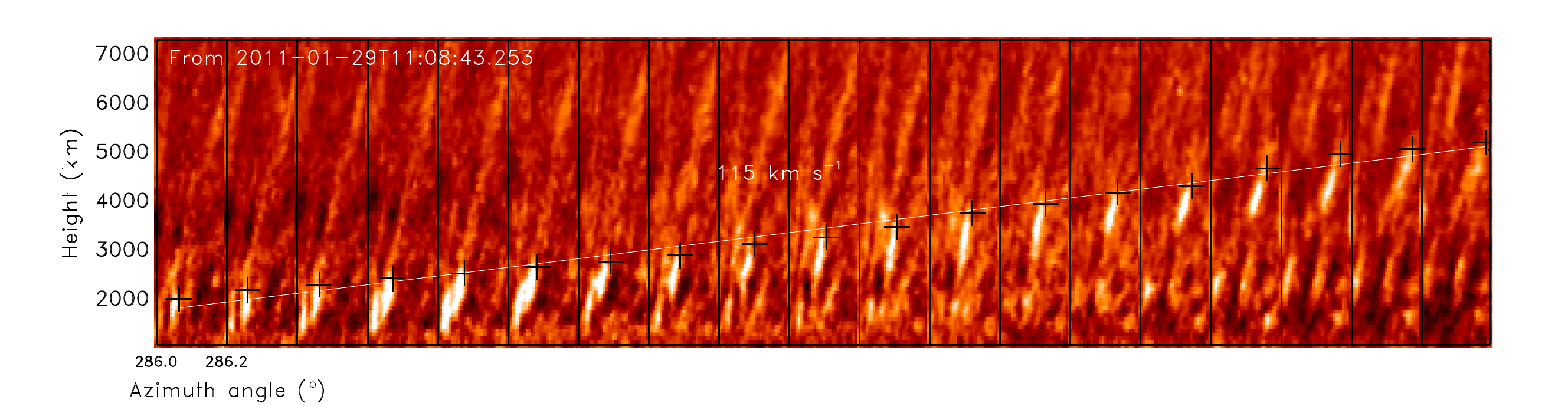}
\includegraphics[scale=0.8, angle=90]{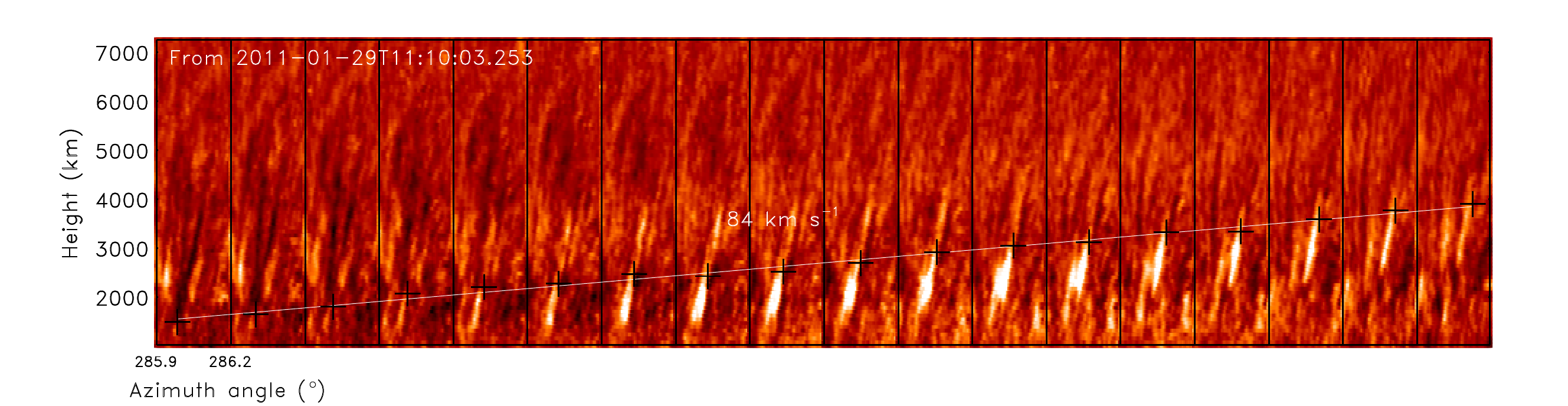}
\includegraphics[scale=0.8, angle=90]{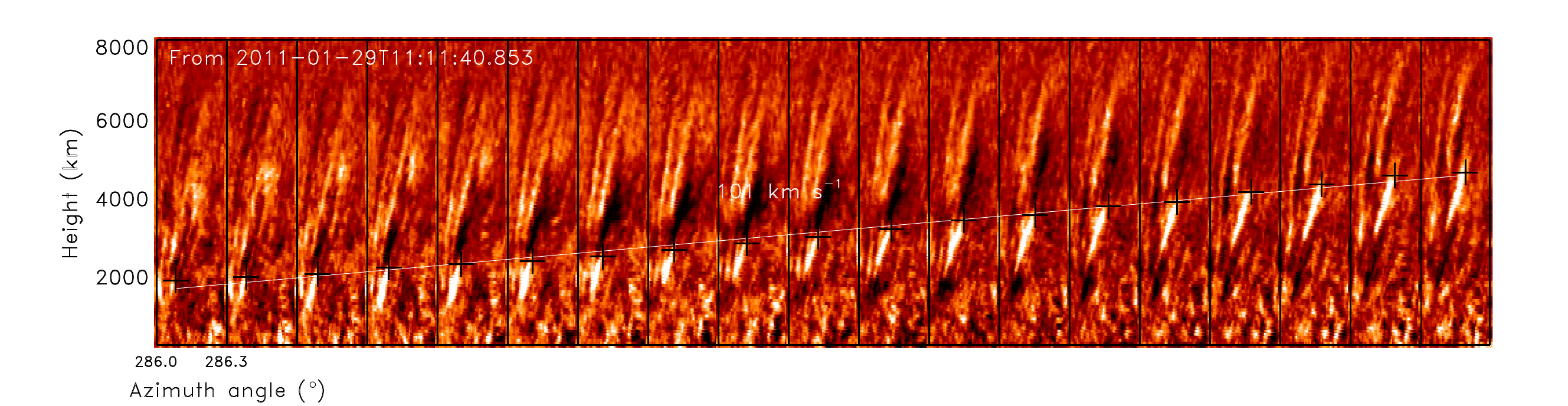}
\caption{
Three examples of the evolution of jets at every 1.6 seconds.
The caption is the same with Figure~\ref{fig2}.
}
\end{sidewaysfigure}

\begin{sidewaysfigure}
\includegraphics[scale=0.8, angle=90]{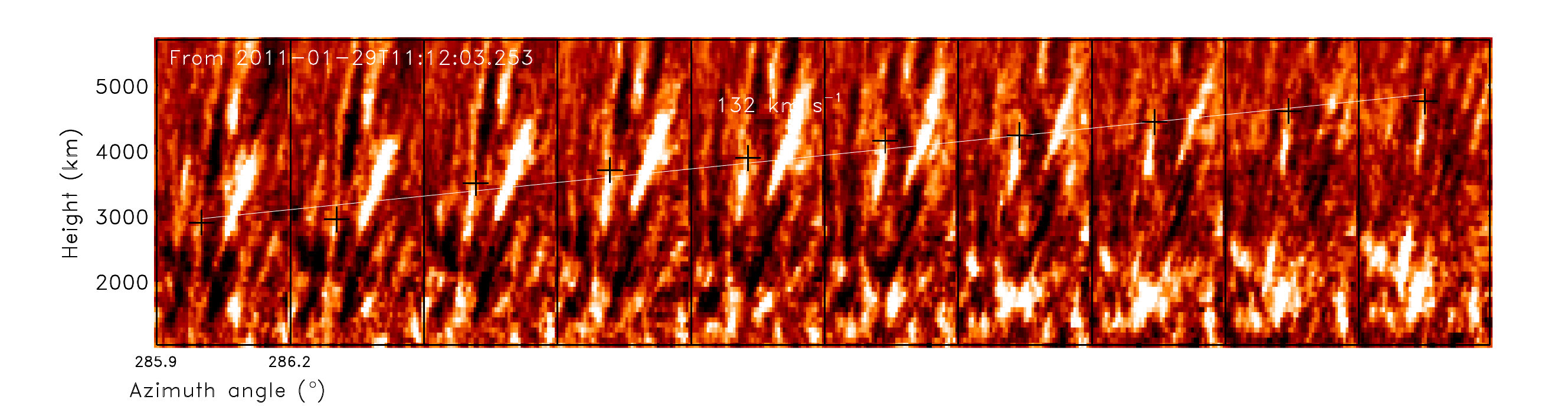}
\includegraphics[scale=0.8, angle=90]{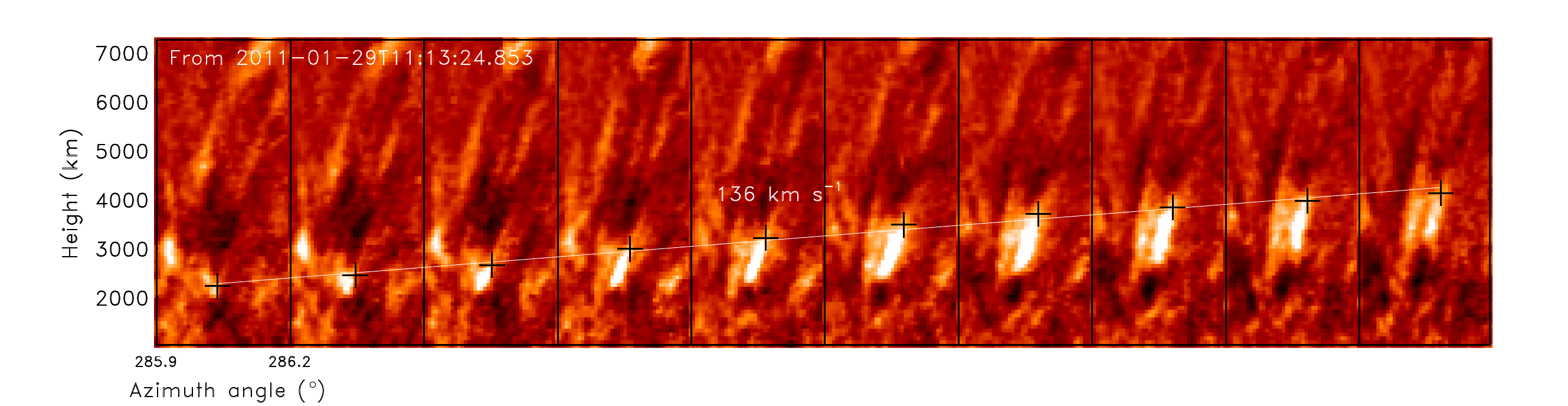}
\includegraphics[scale=0.8, angle=90]{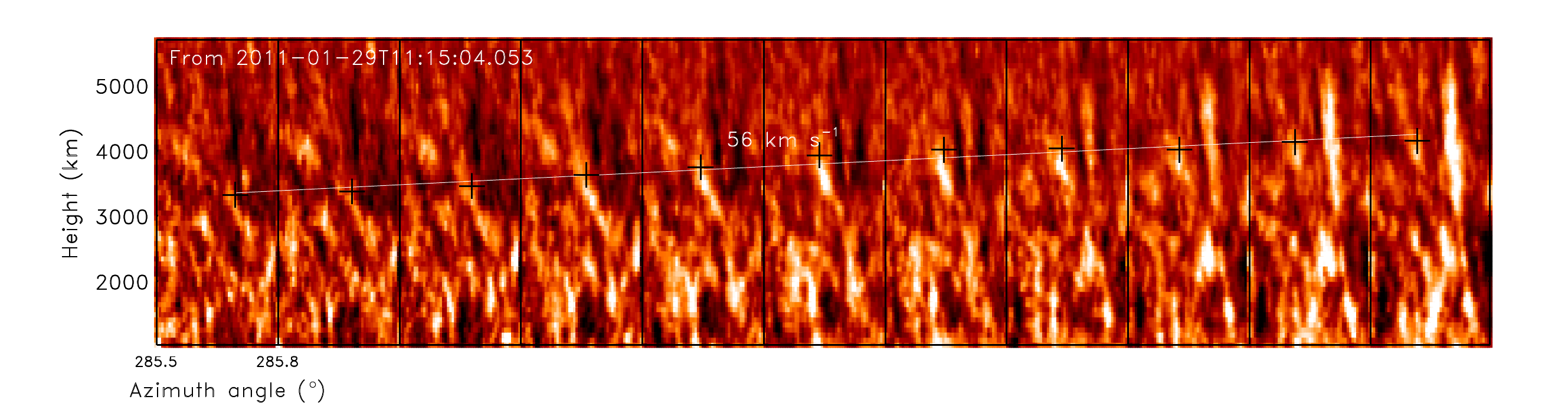}
\caption{
Three examples of the evolution of jets at every 1.6 seconds.
The caption is the same with Figure~\ref{fig2}.
}
\end{sidewaysfigure}

\begin{sidewaysfigure}
\includegraphics[scale=0.8, angle=90]{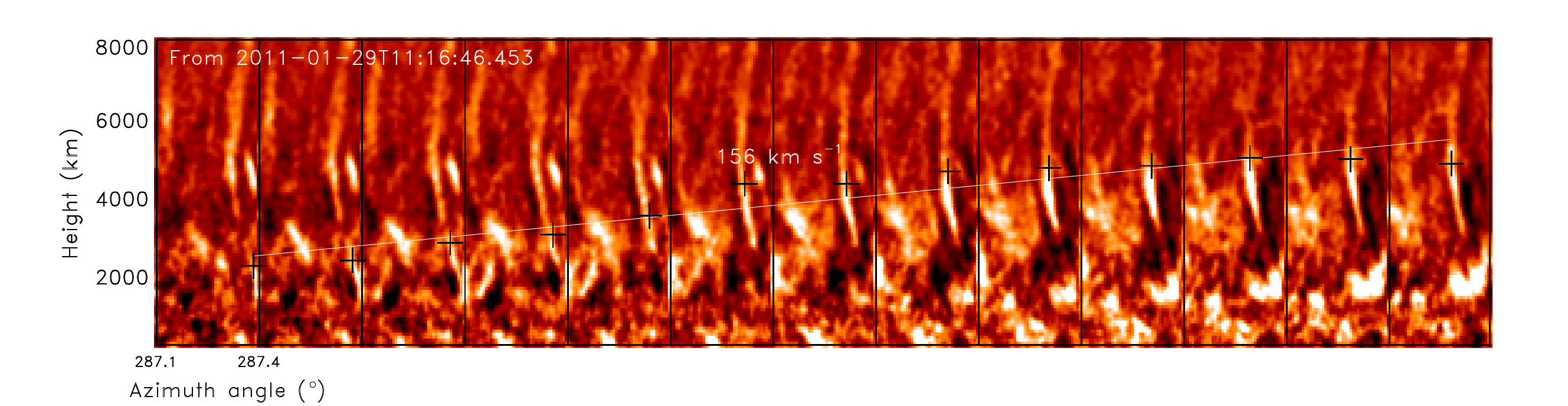}
\includegraphics[scale=0.8, angle=90]{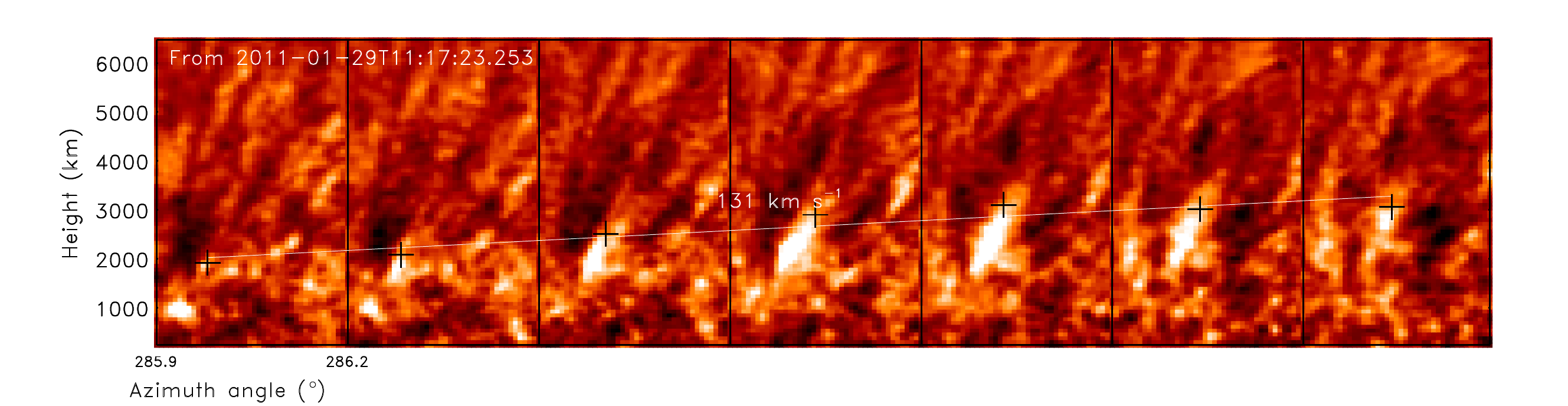}
\includegraphics[scale=0.8, angle=90]{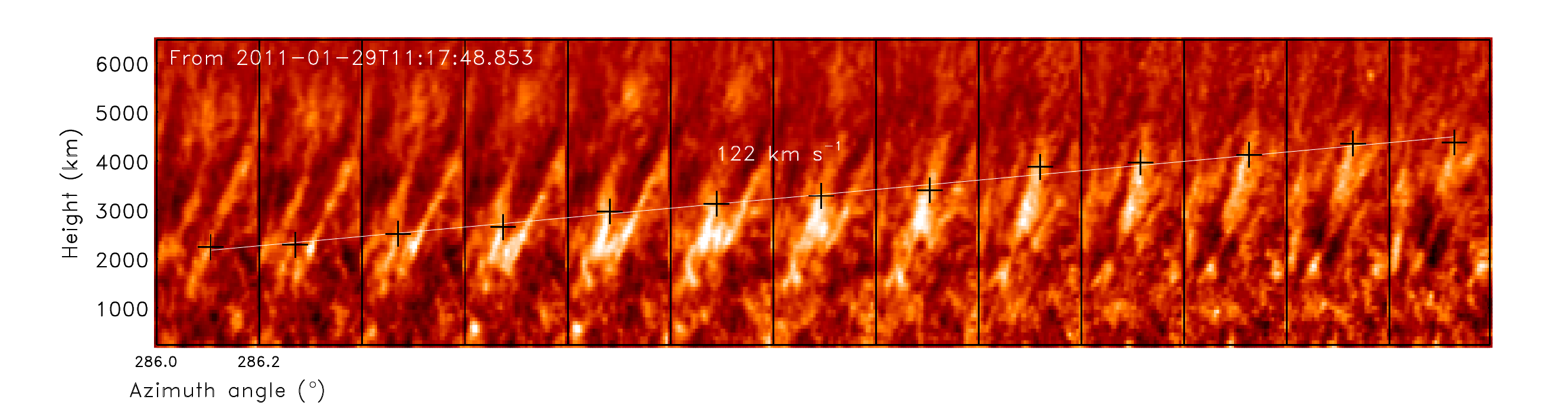}
\caption{
Three examples of the evolution of jets at every 1.6 seconds.
The caption is the same with Figure~\ref{fig2}.
}
\end{sidewaysfigure}

\begin{sidewaysfigure}
\includegraphics[scale=0.8, angle=90]{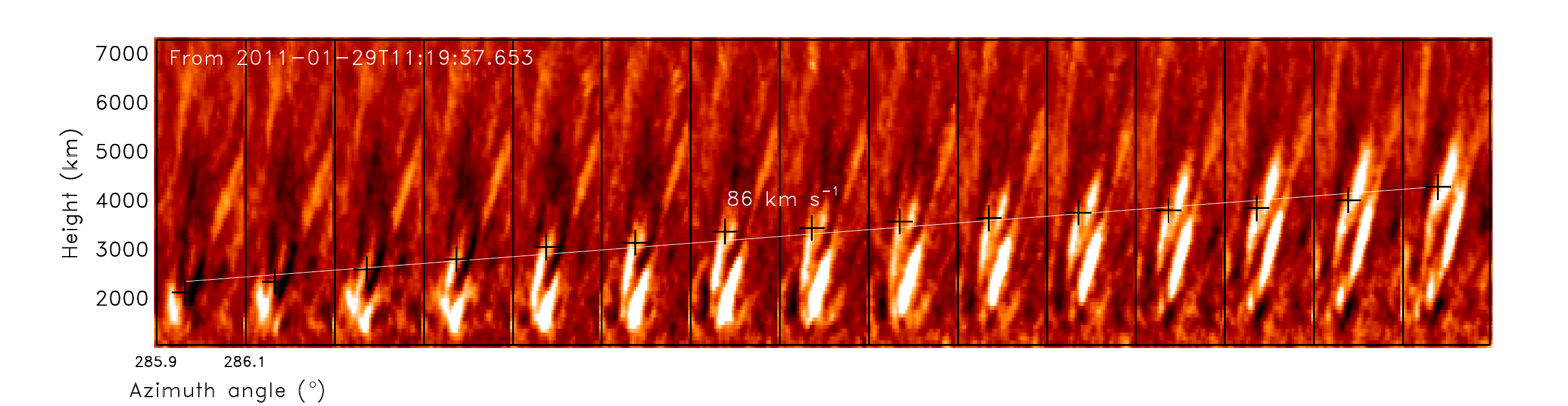}
\includegraphics[scale=0.8, angle=90]{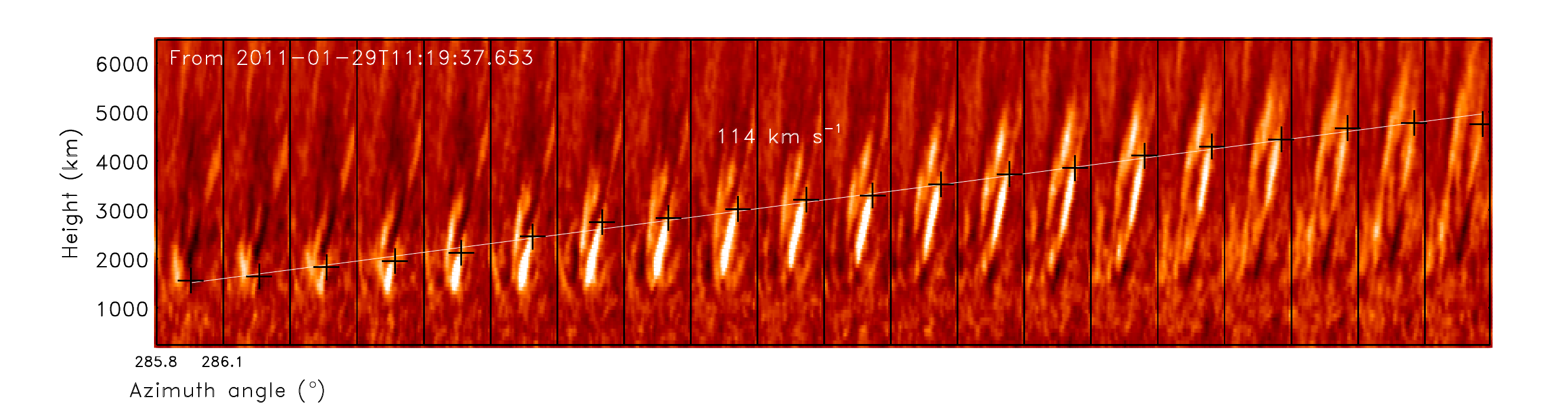}
\includegraphics[scale=0.8, angle=90]{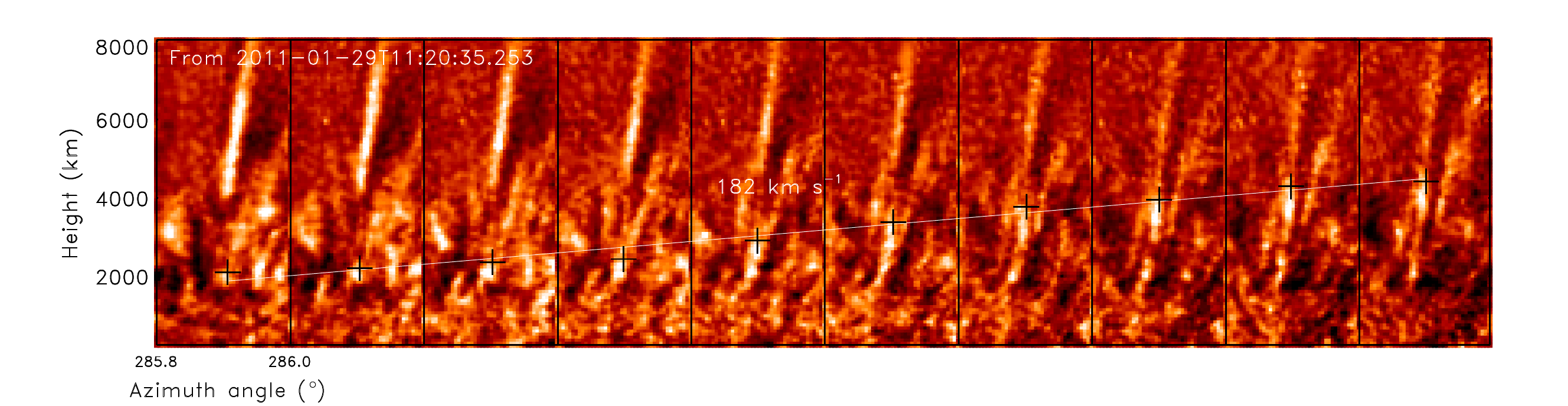}
\caption{
Three examples of the evolution of jets at every 1.6 seconds.
The caption is the same with Figure~\ref{fig2}.
}
\end{sidewaysfigure}

\begin{sidewaysfigure}
\includegraphics[scale=0.8, angle=90]{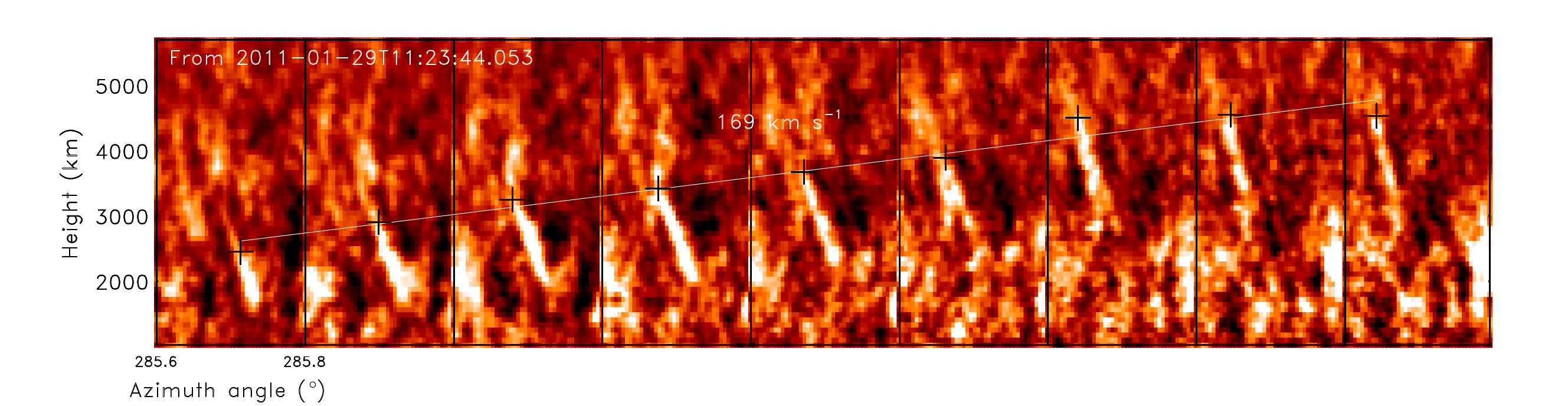}
\includegraphics[scale=0.8, angle=90]{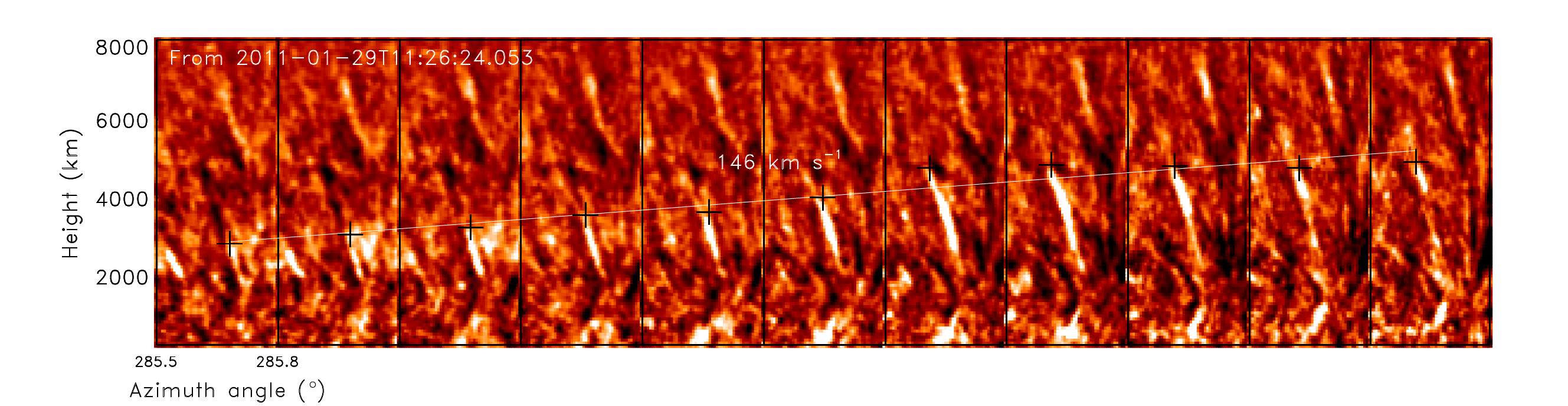}
\includegraphics[scale=0.8, angle=90]{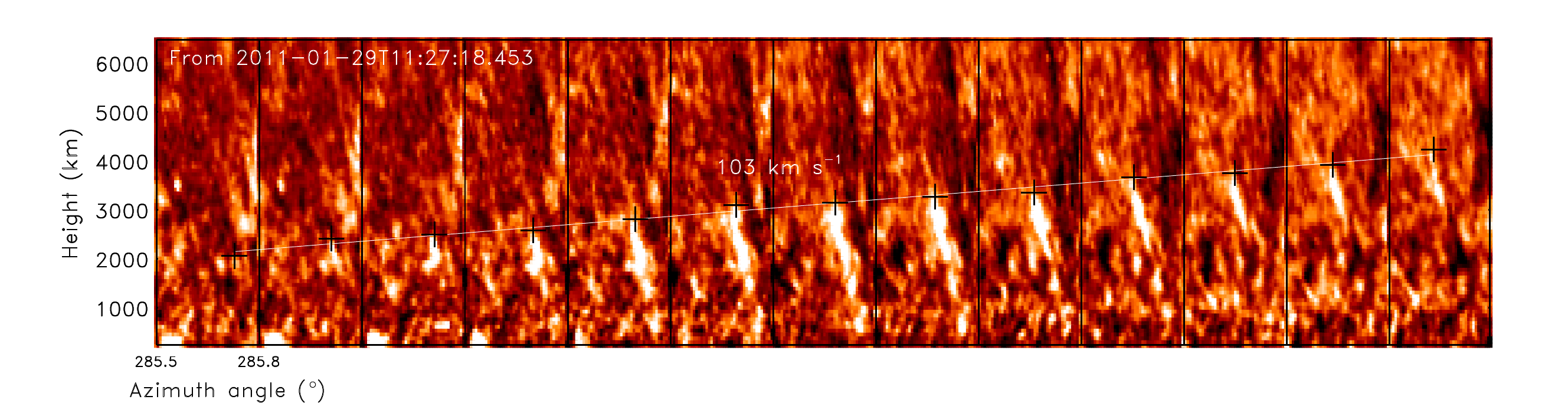}
\caption{
Three examples of the evolution of jets at every 1.6 seconds.
The caption is the same with Figure~\ref{fig2}.
}
\end{sidewaysfigure}

\begin{sidewaysfigure}
\includegraphics[scale=0.8, angle=90]{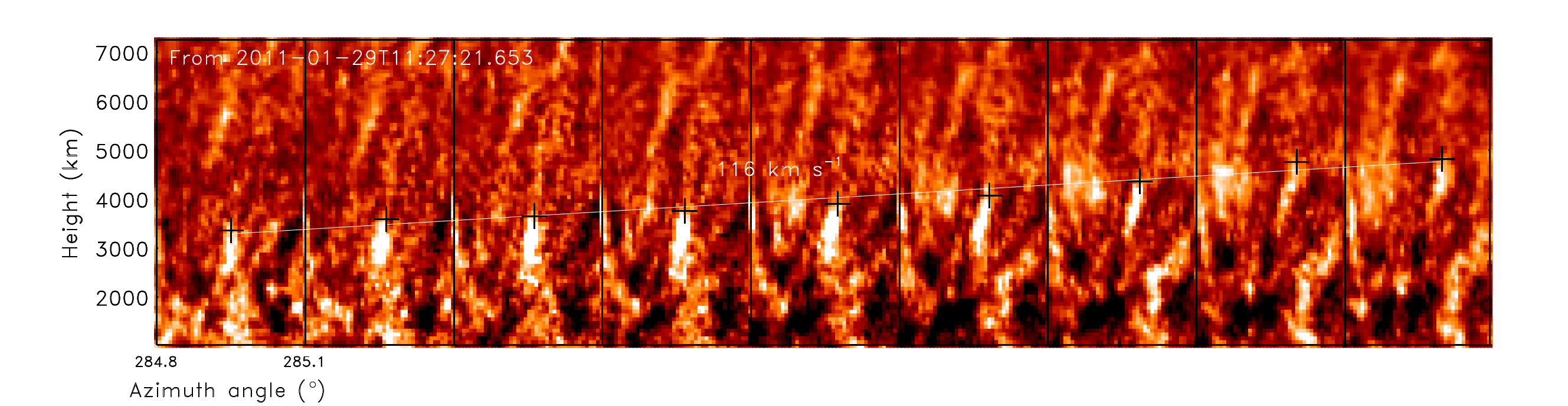}
\includegraphics[scale=0.8, angle=90]{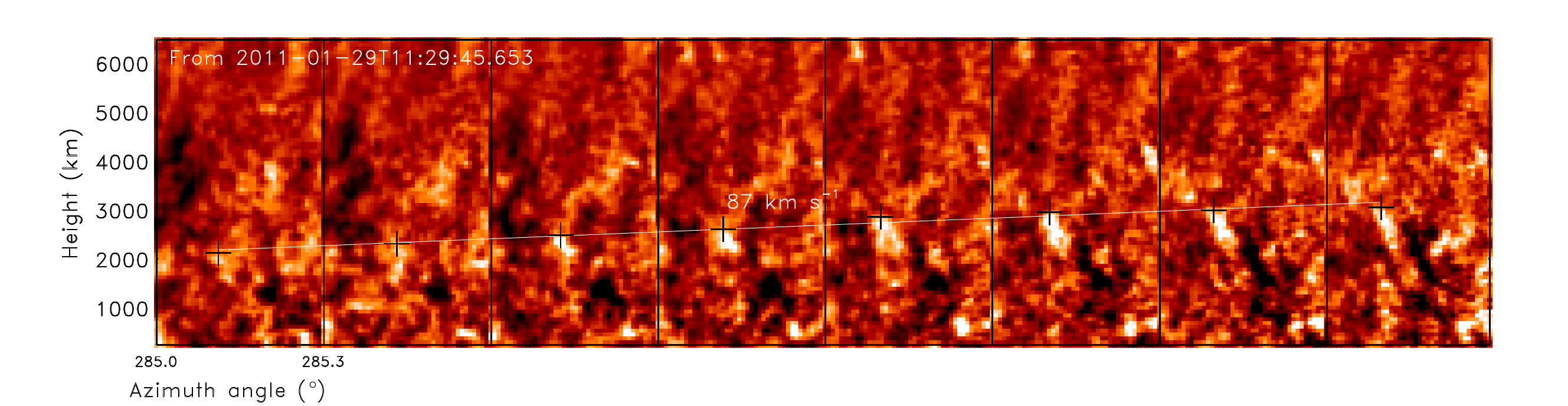}
\includegraphics[scale=0.8, angle=90]{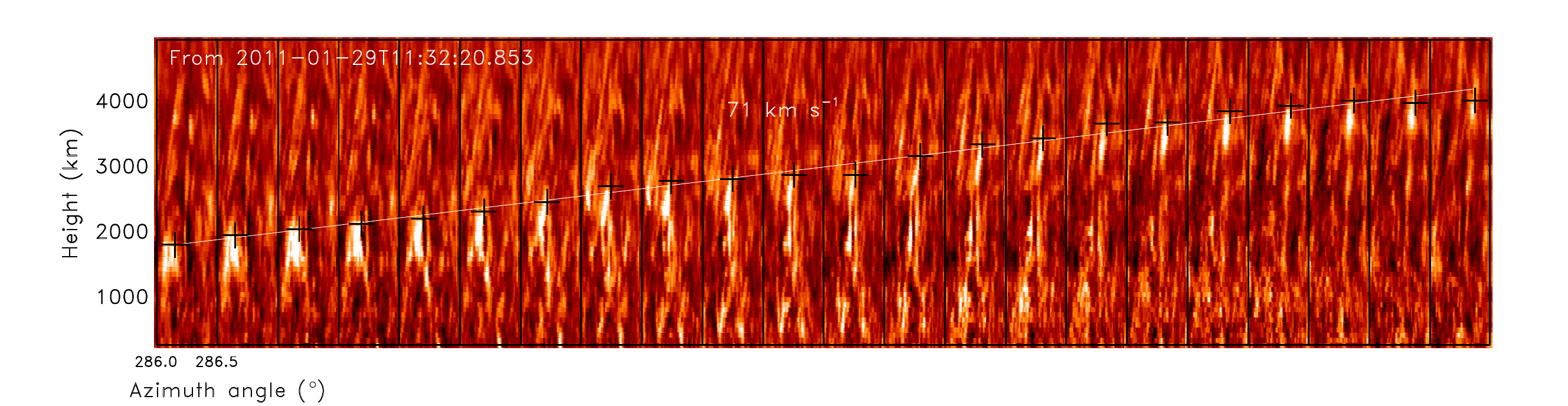}
\caption{
Three examples of the evolution of jets at every 1.6 seconds.
The caption is the same with Figure~\ref{fig2}.
}
\end{sidewaysfigure}

\begin{sidewaysfigure}
\includegraphics[scale=0.8, angle=90]{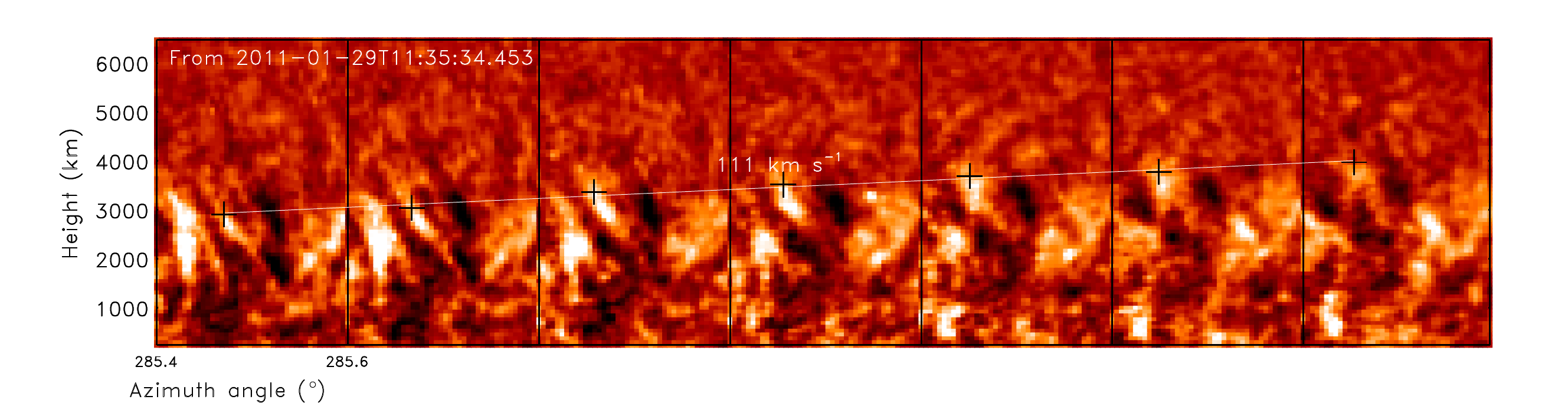}
\includegraphics[scale=0.8, angle=90]{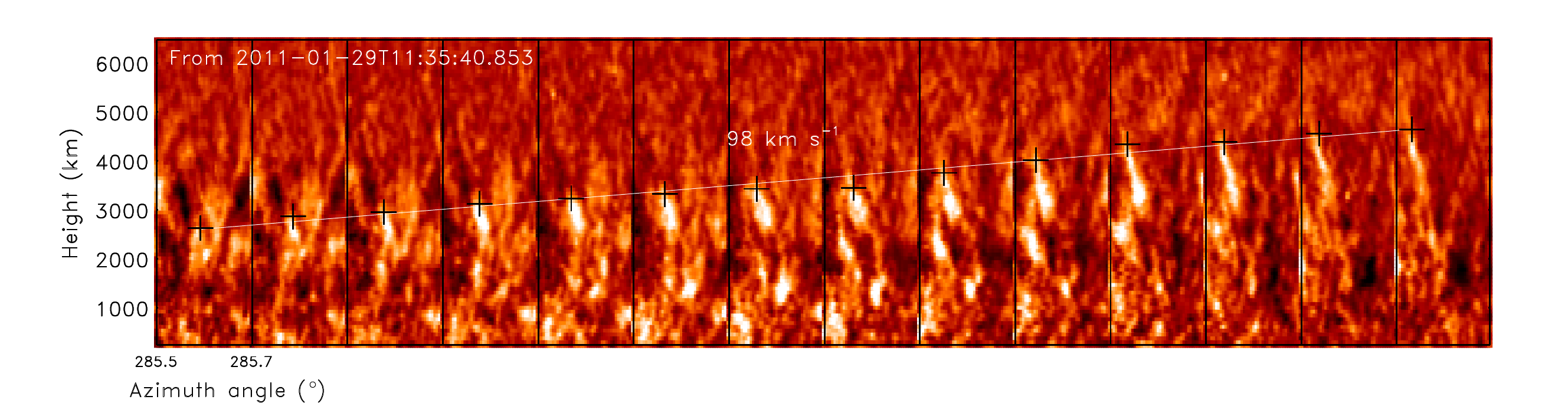}
\includegraphics[scale=0.8, angle=90]{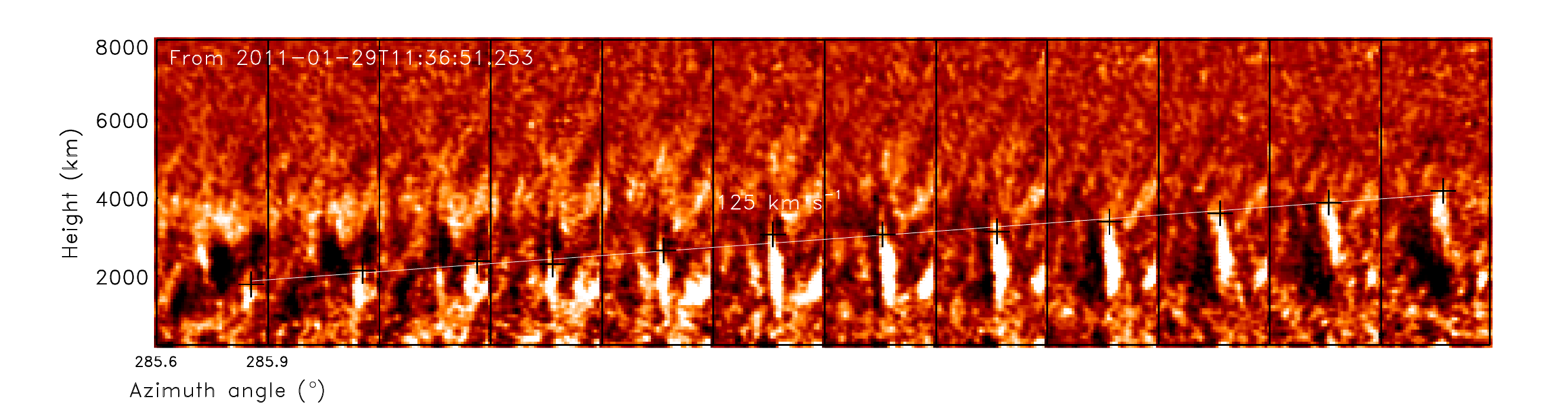}
\caption{
Three examples of the evolution of jets at every 1.6 seconds.
The caption is the same with Figure~\ref{fig2}.
}
\end{sidewaysfigure}

\begin{sidewaysfigure}
\includegraphics[scale=0.8, angle=90]{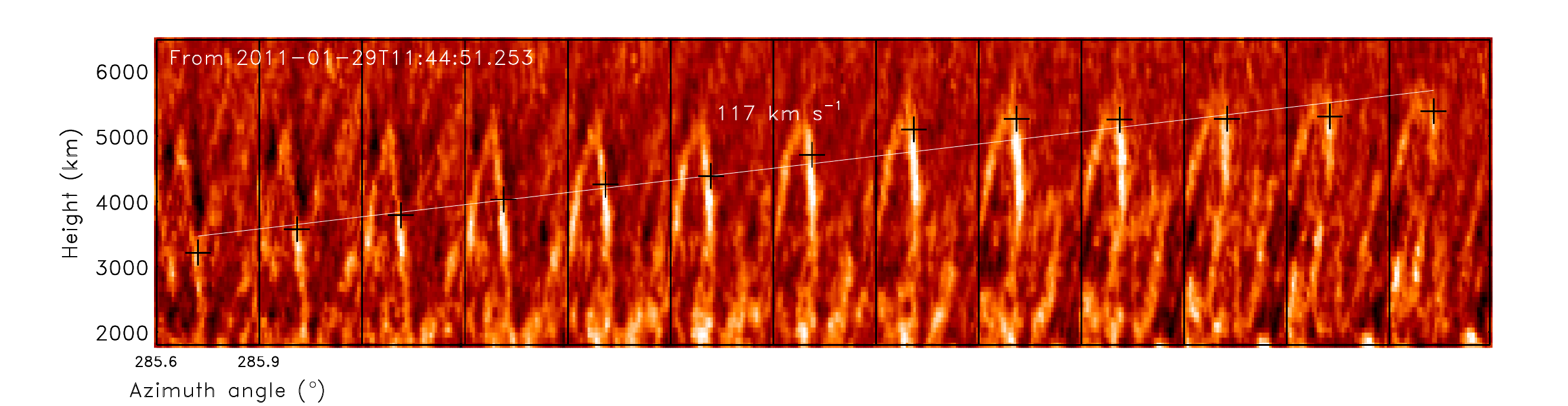}
\includegraphics[scale=0.8, angle=90]{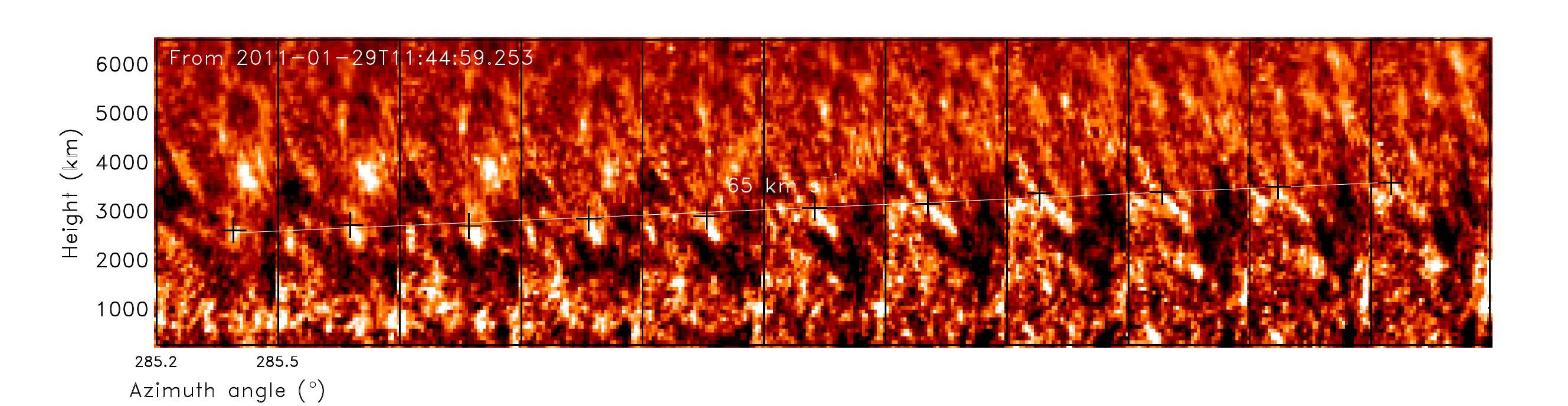}
\includegraphics[scale=0.8, angle=90]{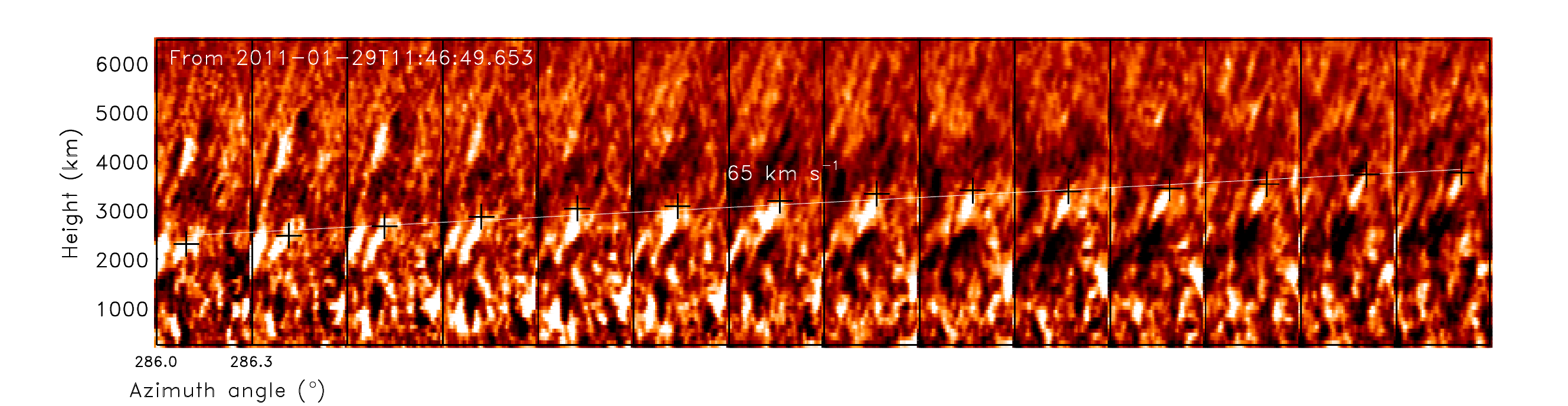}
\caption{
Three examples of the evolution of jets at every 1.6 seconds.
The caption is the same with Figure~\ref{fig2}.
}
\end{sidewaysfigure}

\begin{sidewaysfigure}
\includegraphics[scale=0.8, angle=90]{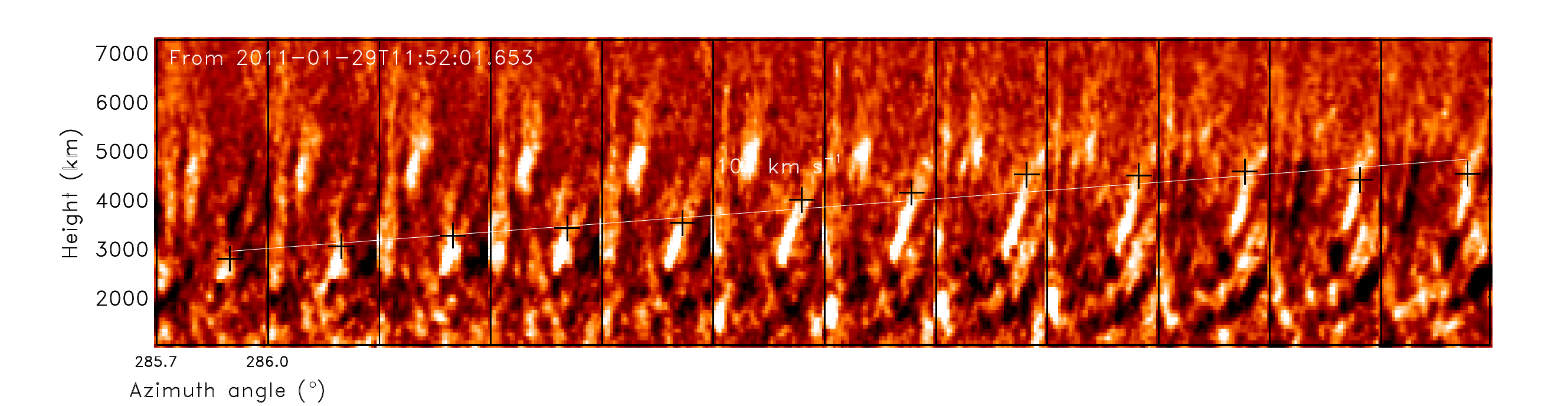}
\includegraphics[scale=0.8, angle=90]{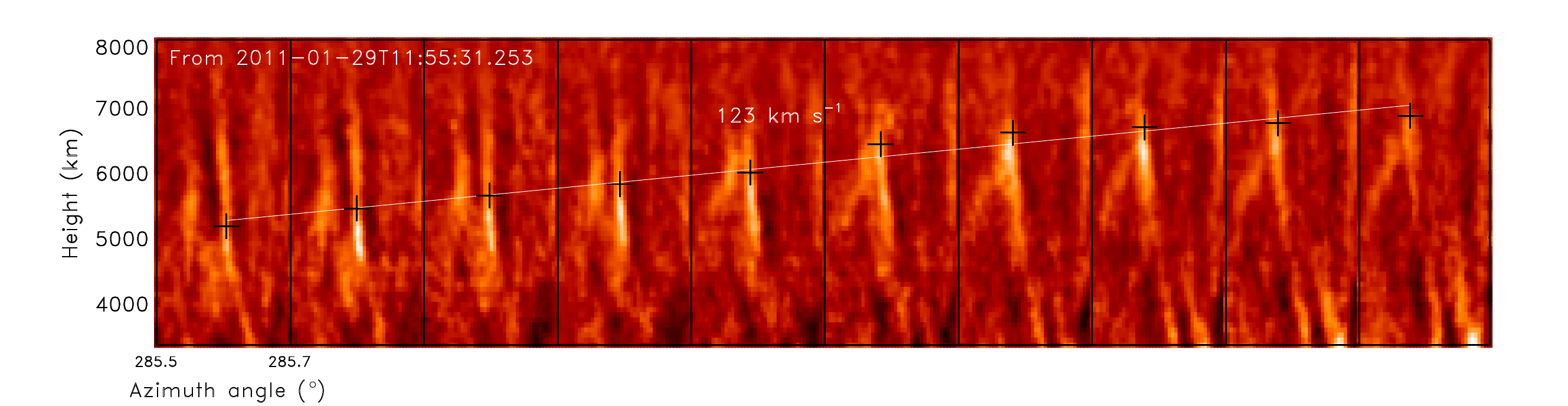}
\includegraphics[scale=0.8, angle=90]{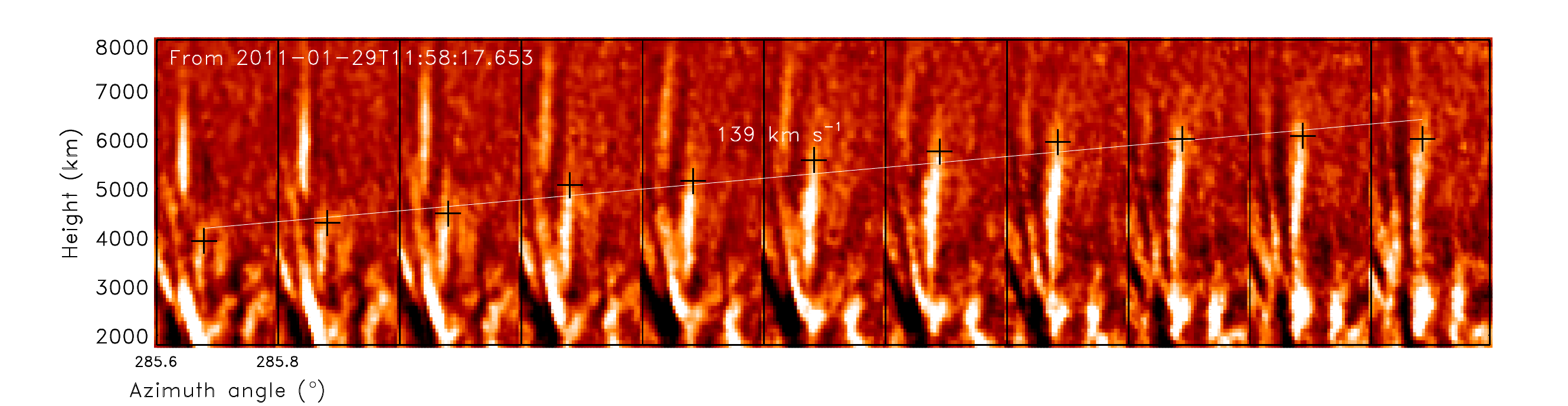}
\caption{
Three examples of the evolution of jets at every 1.6 seconds.
The caption is the same with Figure~\ref{fig2}.
}
\end{sidewaysfigure}

\end{document}